\documentclass{ethpaper}
\usepackage{graphicx}
\usepackage{here}
\usepackage{amssymb}
\usepackage{subfigure}
\usepackage{lineno}

\begin{document}
\begin{titlepage}
\ethnote{}
\title{A visualization of the damage in Lead Tungstate calorimeter crystals\\after exposure to high-energy hadrons}
\begin{Authlist}
G. Dissertori, D.~Luckey, F.~Nessi-Tedaldi, F.~Pauss, R.~Wallny
\Instfoot{eth}{Institute for Particle Physics, ETH Zurich, 8093 Zurich, Switzerland}
R. Spikings, R. Van der Lelij
\Instfoot{unige}{Department of Mineralogy, University of Geneva, 1205 Geneva  4, Switzerland}
G.~Arnau~Izquierdo
\Instfoot{cern}{CERN - EN Department, 1211 Geneva 23, Switzerland}
\end{Authlist}
\maketitle
\begin{abstract}
The anticipated performance of calorimeter crystals in the environment expected after the planned High-Luminosity upgrade of the Large Hadron Collider (HL-LHC) at CERN has to be well understood, before informed decisions can be made on the need for detector upgrades. Throughout the years of running at the HL-LHC, the detectors will be exposed to considerable fluences of fast hadrons, that have been shown to cause cumulative transparency losses in Lead Tungstate scintillating crystals. In this study, we present direct evidence of the main underlying damage mechanism. Results are shown from a test that yields a direct insight into the nature of the hadron-specific damage in Lead Tungstate calorimeter crystals exposed to 24 GeV/c protons.
\end{abstract}
\vspace{7cm}
\conference{submitted to Elsevier for publication in Nucl. Instr. and Meth. in Phys. Research A}
\end{titlepage}
\section{Introduction}
\label{s-INT}
Detectors now in operation at the Large Hadron Collider at CERN may need to
face an increasingly challenging environment after the accelerator upgrade to
High-Luminosity running (HL-LHC), that is being planned to start - according to the present schedule - in 2022.
While performance requirements will be driven by the physics results obtained at the LHC in the near future,
a thorough understanding of the detector behavior is needed, as an input to knowledgeable decisions
concerning upgrades.

In this framework, our 
earlier studies on Lead Tungstate
(PbWO$_4$)~\cite{r-LTNIM,r-pionNIM,r-LYNIM}  have established that hadronic showers from
high-energy
protons~\cite{r-LTNIM} and pions~\cite{r-pionNIM} cause a
cumulative loss of Light Transmission in PbWO$_4$, which is permanent at room temperature,
while  no hadron-specific
change in scintillation emission~\cite{r-LYNIM} was seen. 
The features observed 
hint at 
local centers of damage that might be caused by 
fragments of the heavy elements, Pb and W as the dominant cause of transmission losses.
Such fragments
can have a range up to 10 $\mu$m and energies up to $\sim$100 MeV,
corresponding to a stopping power four orders of magnitude higher than the one of 
minimum-ionizing particles~\cite{r-LTNIM}.

The qualitative understanding we gained of hadron damage in Lead
Tung\-state led us to predict~\cite{r-CAL08} that such hadron-specific
damage contributions are absent in crystals  consisting only of
elements with $Z < 71$, which is the experimentally observed threshold
for fission~\cite{r-THR}, while they should be expected in crystals
containing elements with  $Z > 71$. We confirmed the first prediction with measurements~\cite{r-CEF3}
that show how hadrons in Cerium Fluoride cause a damage that recovers at room temperature, 
with none of the features we observed for Lead Tungstate.
The second prediction is confirmed by existing proton-damage measurements
in BGO~\cite{r-KOBA,r-DPF}, in Lead Fluoride and BSO 
~\cite{r-KOR}, which all contain elements with $Z > 71$.

All this indirect evidence is consistent with a mechanism by which the heavy fission fragments
deposit a lot of energy along their short track, leaving regions within the crystal where the lattice
structure is modified: it can remain disturbed, strained, disordered, or re-oriented.
These damage regions have different optical and mechanical properties from the surrounding 
crystal lattice, and thus they can act as scatterers for light propagating in the crystal, whatever its origin,
from scintillation or from an external source.

The mechanism is well known since a long time, and is commonly called "fission track damage"
in literature. In fact, almost a century ago, F. Dessauer formulated the concept of a "thermal
spike" when an incident ion comes to a stop in matter~\cite{r-DES}. Later on, L.~T.~ Chadderton and I. McC. Torrens , in
their book on the subject~\cite{r-CHA} , explained how 
{\em ``Along the heated cylindrical track of the fragment the crystalline matter is disturbed,
decomposed, or removed. The subsequent arrangement is not necessarily perfect and strain
centers or dislocations remain''} and ion implantation has become, in recent days, a technique used
to change material properties~\cite{r-TOW}.

In this study we present direct, visual
evidence that confirms our understanding of hadron damage mechanisms. This will help making
informed decisions regarding future calorimeter design or upgrades, to ensure an optimum long-term
detector performance.

\section{Macroscopic observations}
\label{s-MAC}
\begin{figure}[b]
\begin{center}
{\mbox{\includegraphics[width=12cm]{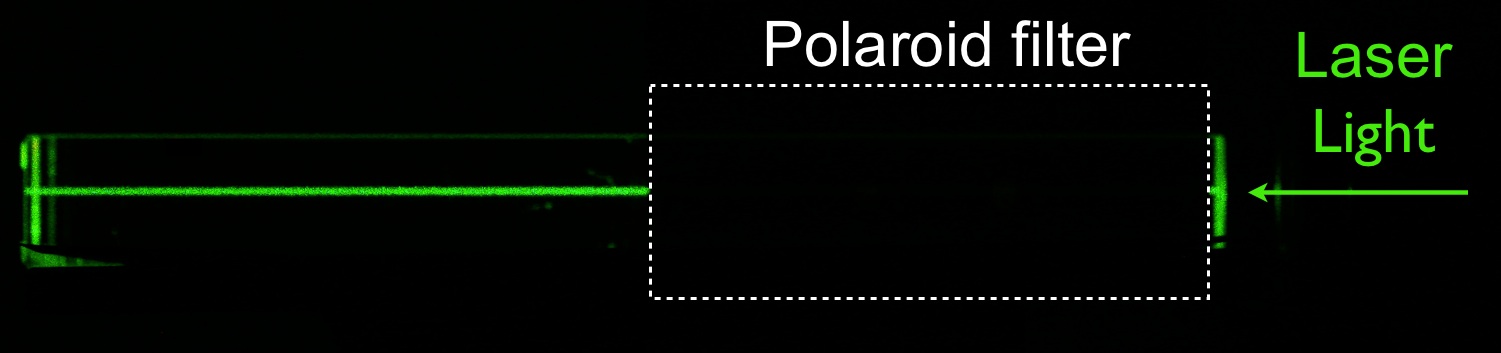}}}
\end{center}
\caption{Laser light (here at 543.5 nm) is scattered, and thus becomes visible, when shone through a
hadron-irradiated Lead Tungstate crystal. A polaroid filter 
(indicated by the dashed line) blocks the scattered light, revealing its linear polarization.}\label{f-LAS}
\end{figure}
The features of hadron-induced changes in Light Transmission published
in~\cite{r-LTNIM} are the peculiar ones of Rayleigh scattering, which implies 
the presence of dipole-shaped regions acting as
scatterers for the light. When green 543.5 nm LASER light is shone through a
hadron-irradiated crystal, as exemplified in the photograph of Fig.~\ref{f-LAS}, its light gets
scattered, making the beam visible. In the same photograph,
a Polaroid filter was introduced between the observer and the crystal, as indicated by the 
dashed rectangle, so as to cover the right half of the crystal: the scattered LASER light
is not transmitted by the filter. This demonstrates how the scattered light is polarized,
yet another feature that is distinctive for
Rayleigh scattering, and thus for the presence of localized regions of damage acting as
scatterers for the light. All these observations were also
observed at different wavelengths, as verified with 633 nm LASER light.

\section{Microscopy observations - the method}
\label{s-MIC}
While the macroscopic observations in Section~\ref{s-MAC} reveal the presence of
scatterers, these are not individually visible to the naked eye. It was thus interesting to 
pursue a microscopy visualization, following the fission track analysis method used
 in geochronology for mineral dating~\cite{r-FLE}. The method is commonly used on 
minerals containing Uranium, which naturally exhibit damage tracks caused 
by the daughter products from spontaneous fission of $^{238}\mathrm{U}$. Geochronological mineral  dating 
relies on the count of these fission tracks, but it also needs a complementary 
information: the Uranium concentration at a given time. Based on the 
${^{235}{\mathrm U}}/{^{238}{\mathrm U}}$ abundance ratio, the Uranium concentration can be
determined by inducing $^{235}\mathrm{U}$ fission in the studied mineral through thermal neutron irradiation,
with an external detector, such as high-purity muscovite mica, held in intimate
contact with a polished section of the mineral surface.
The $^{235}{\mathrm U}$  fission fragments create induced tracks that reach the overlying external
detector, where they are later revealed by chemical etching to make them easily visible
for counting under an optical microscope and thus infer the $^{238}\mathrm{U}$ concentration. 
It was thus obvious to apply this well-established methodology to attempt revealing induced
fission damage in Lead Tungstate.

\section{The irradiation setup}
\label{s-SUP}
We have performed hadron irradiations using a setup suitable for allowing the visualization of
damage trails, using  the 24\,GeV/c proton beam at the IRRAD1
facility\,\cite{r-IR1} located in the T7 beam line of the CERN PS accelerator.
\begin{figure}[b]
\begin{center}
{\mbox{\includegraphics[width=8cm]{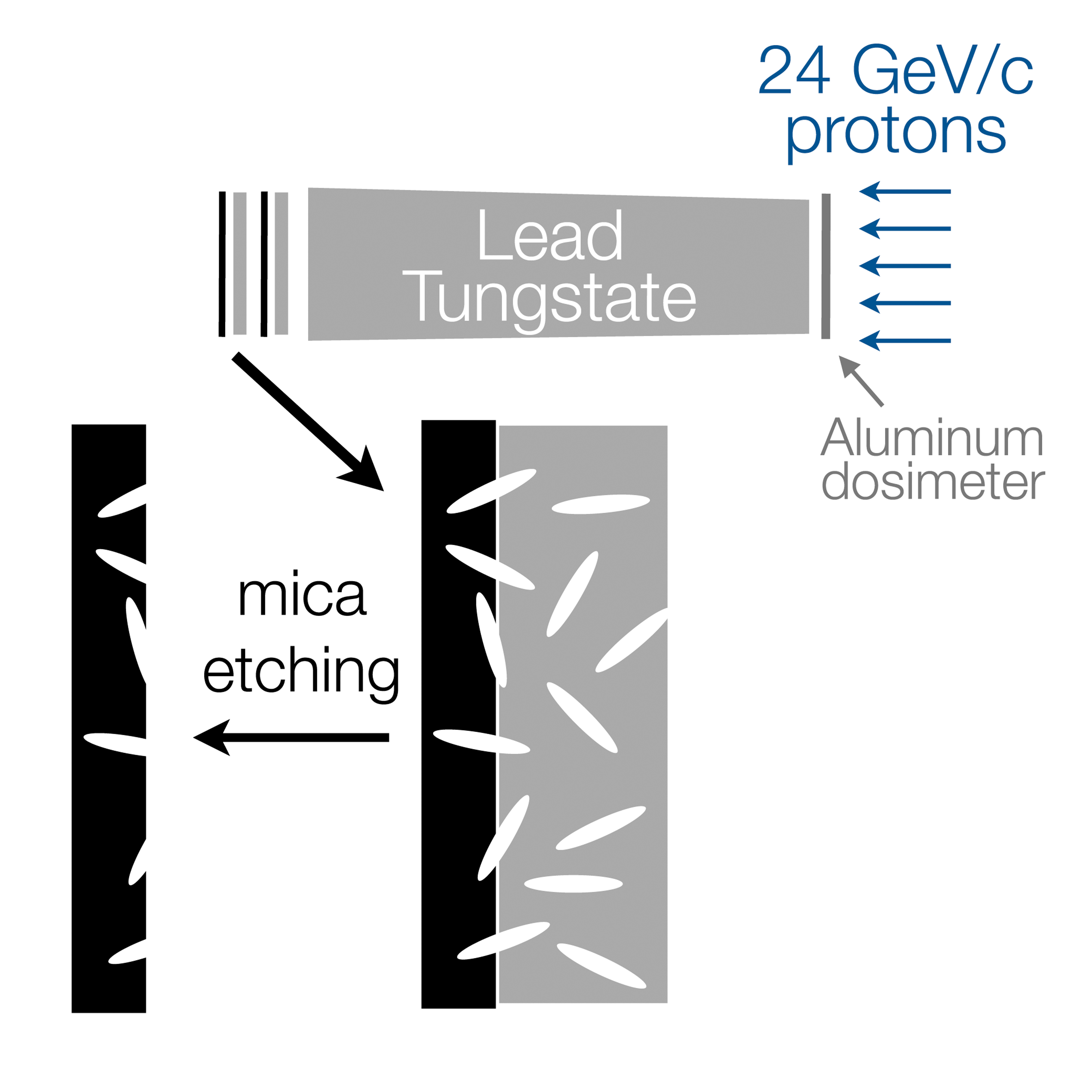}}}
\end{center}
\caption{Irradiation setup for fission-track visualization (see Sec.~\ref{s-SUP} for a detailed description)}\label{f-SUP}
\end{figure}

For each irradiation, two samples have been prepared, where
a Lead Tungstate slide, 2 mm thick and 1~cm $\times$ 2~cm in cross section,
has been placed in intimate contact with a high-purity
muscovite mica ($\mathrm{KAl}_2\mathrm{(AlSi}_3\mathrm{O}_{10}\mathrm{)(F,OH)}_2$)
slide of same cross section, 0.1 mm thick. The contact was ensured by a 
heat-shrinkable plastic wrapping. The samples have been placed behind a 7.5 cm long Lead
Tungstate crystal for the irradiation, to make sure the running conditions are reproduced,
that are encountered in homogeneous high-energy physics calorimetry, where energetic hadrons
cause penetrating hadron showers, i.e. a cascade of nuclear interactions which are at the
origin of the observed hadron-specific damage.
The schematics of the irradiation setup is visible in Fig.~\ref{f-SUP}.
For the hadron irradiation, the proton beam was broadened to cover the whole crystal cross section, and the fluence for
each irradiation was determined following the method described in~\cite{r-LTNIM},
from the activation of a high-purity aluminum foil covering the crystal front face.

Three irradiations were performed, on different pairs of
samples, up to fluences of, respectively,
 $\Phi^1_p=(1.17 \pm 0.09) \times
10^{11}\;\mathrm{cm^{-2}}$, $\Phi^2_p=(1.59 \pm 0.13) \times
10^{12}\;\mathrm{cm^{-2}}$  and $\Phi^3_p=(0.73 \pm 0.09) \times
10^{13}\;\mathrm{cm^{-2}}$.

\section{Visualization of damage tracks in mica}
\label{s-irrad}
\begin{figure}[H]
\begin{center}
{\mbox{\includegraphics[width=12cm]{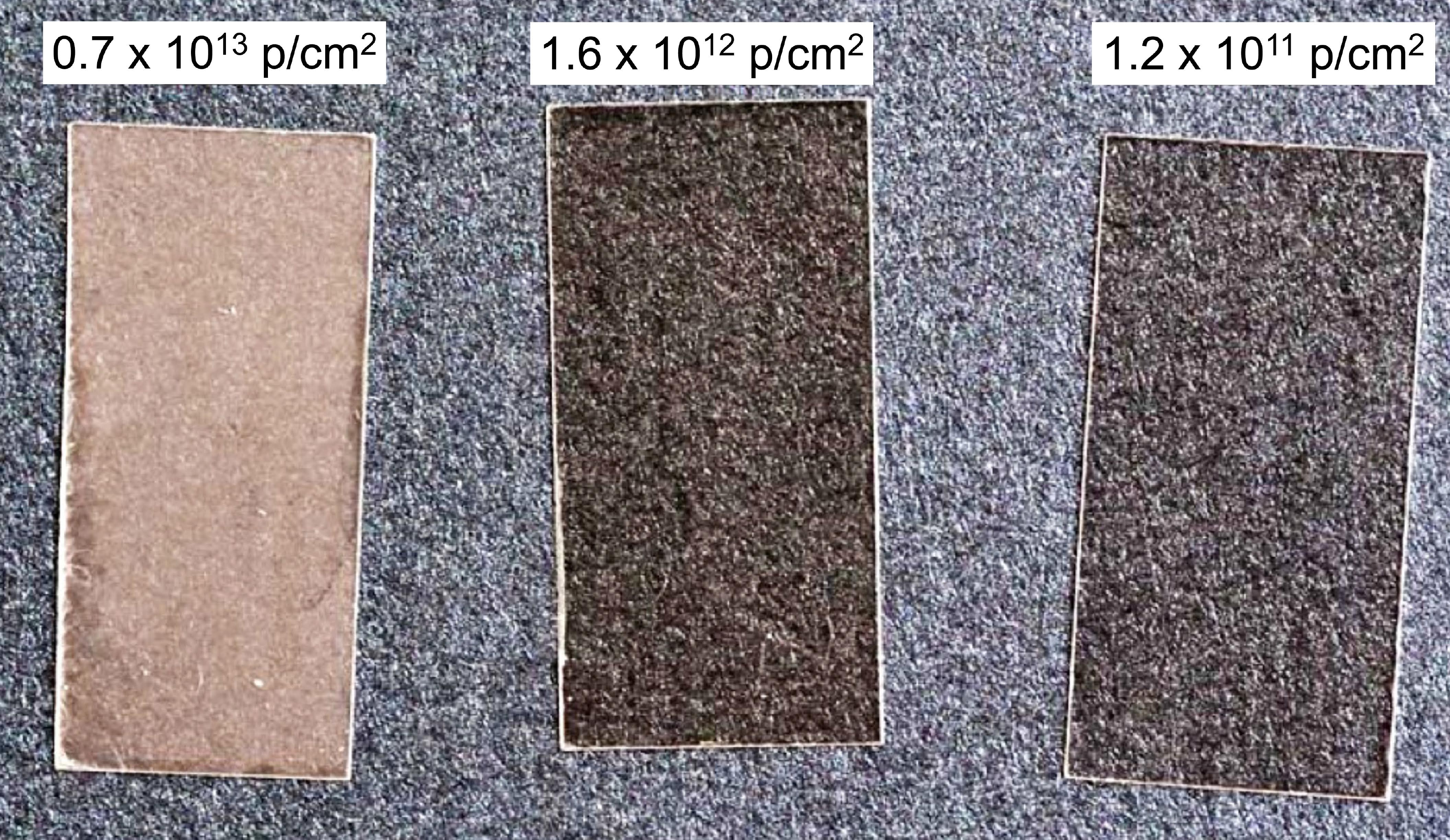}}}
\end{center}
\caption{Mica slides, after chemical etching,
that were irradiated up to three different fluences in intimate contact with Lead Tungstate, as an
external detector for fission tracks.}\label{f-ALL}
\end{figure}

Fission damage in mica
changes its crystalline structure into a metamict state, that
is easily etched according to a well-known recipe~\cite{r-ETC},
whereby the slides
are immersed in HF at 40\% for 45 min at $20^o$ C. The procedure
allows to remove the material of the damaged region, leaving the
surrounding crystalline matter intact.
Examining by eye the mica slides after chemical etching (Fig.~\ref{f-ALL}),
one notices a coloration that intuitively correlates with the
applied irradiation fluence.

\begin{figure}[H]
\begin{center}
  \begin{tabular}[h]{c}
{\mbox{\includegraphics[width=70mm]{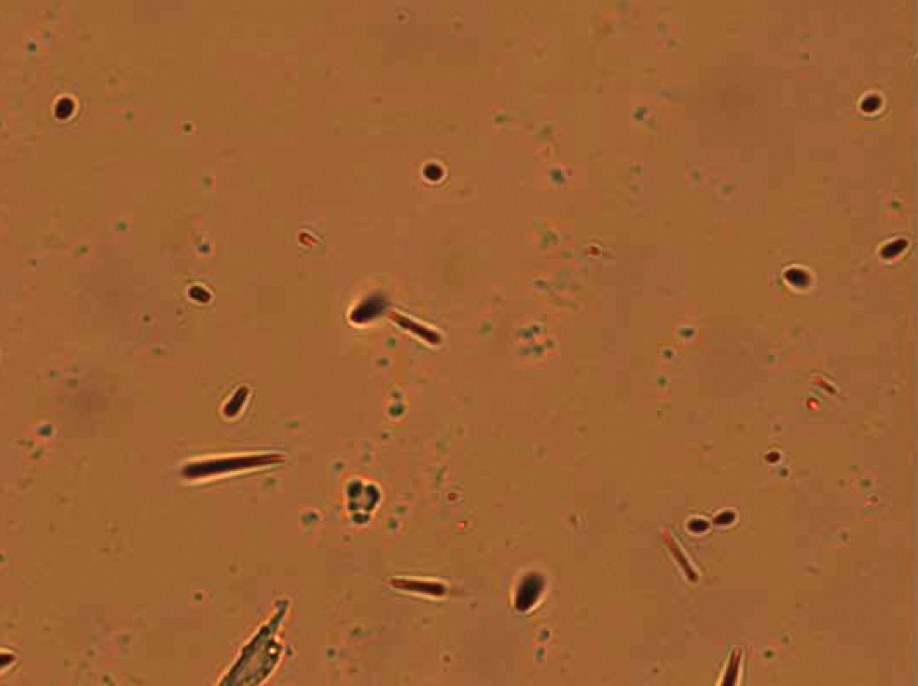}}}\\
{\mbox{\includegraphics[width=70mm]{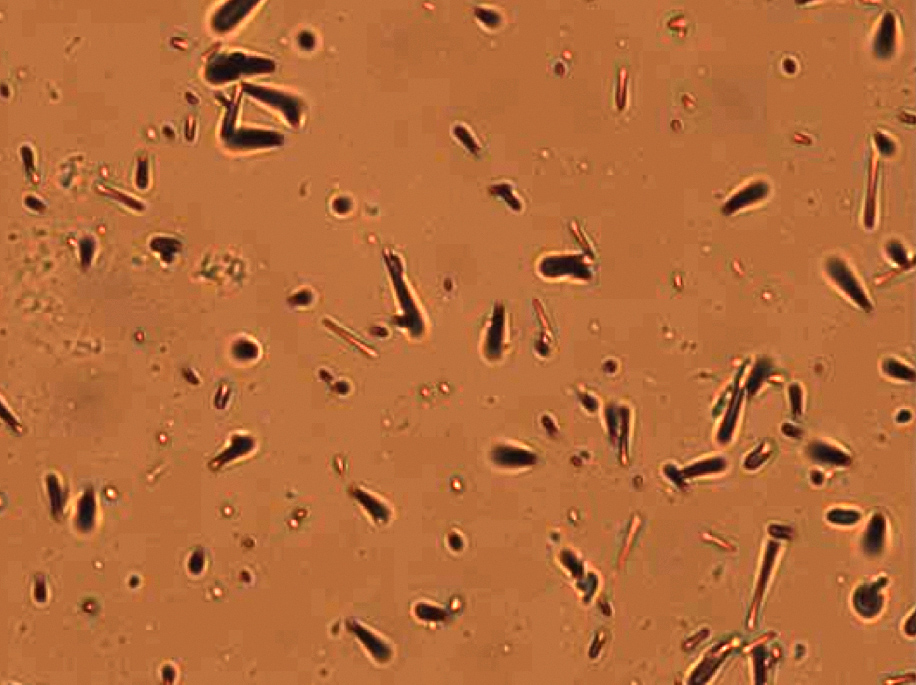}}}\\
{\mbox{\includegraphics[width=70mm]{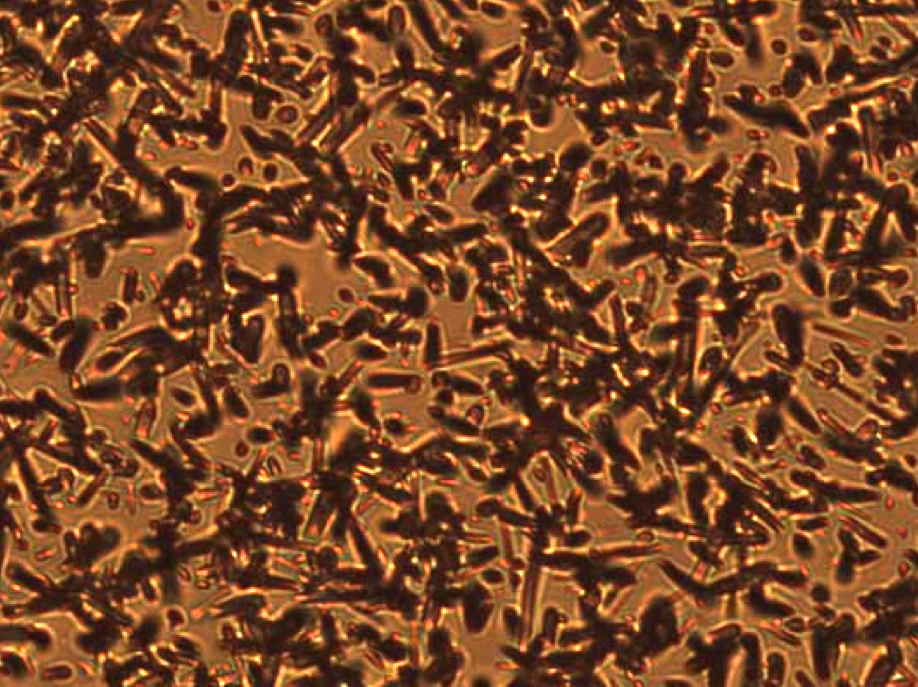}}}
   \end{tabular}
\end{center}
\caption{Transmitted-light optical-microscopy images of fission tracks in the mica slides from Fig.~\ref{f-ALL}
 for the three irradiation fluences,
$\Phi^1_p$ (top), $\Phi^2_p$ (middle) and $\Phi^3_p$ (bottom).\label{f-TRA}}
\end{figure}

The mica slides were then examined with an Axio Imager Z1M
optical microscope from Zeiss~\cite{r-ZEI}. Transmitted light images are shown in Fig.~\ref{f-TRA}.
Linear damage tracks are visible, and they were counted at 1000x magnification but at the highest fluence,
where individual tracks couldn't be visually resolved. Their surface densities clearly
correlate with the irradiation fluences, as listed in
Table~\ref{t-TB1}. Even where, for the highest fluence, no
accurate count was possible, the proportionality is visually striking.

It can be qualitatively observed how the 
damage tracks are straight and occur in random orientations, as expected due to fission 
fragments produced by interactions in a hadron shower, and that they do not follow any
pre-existing fabric in the mica.

\begin{figure}[H]
\begin{center}
  \begin{tabular}[h]{c}
{\mbox{\includegraphics[width=70mm]{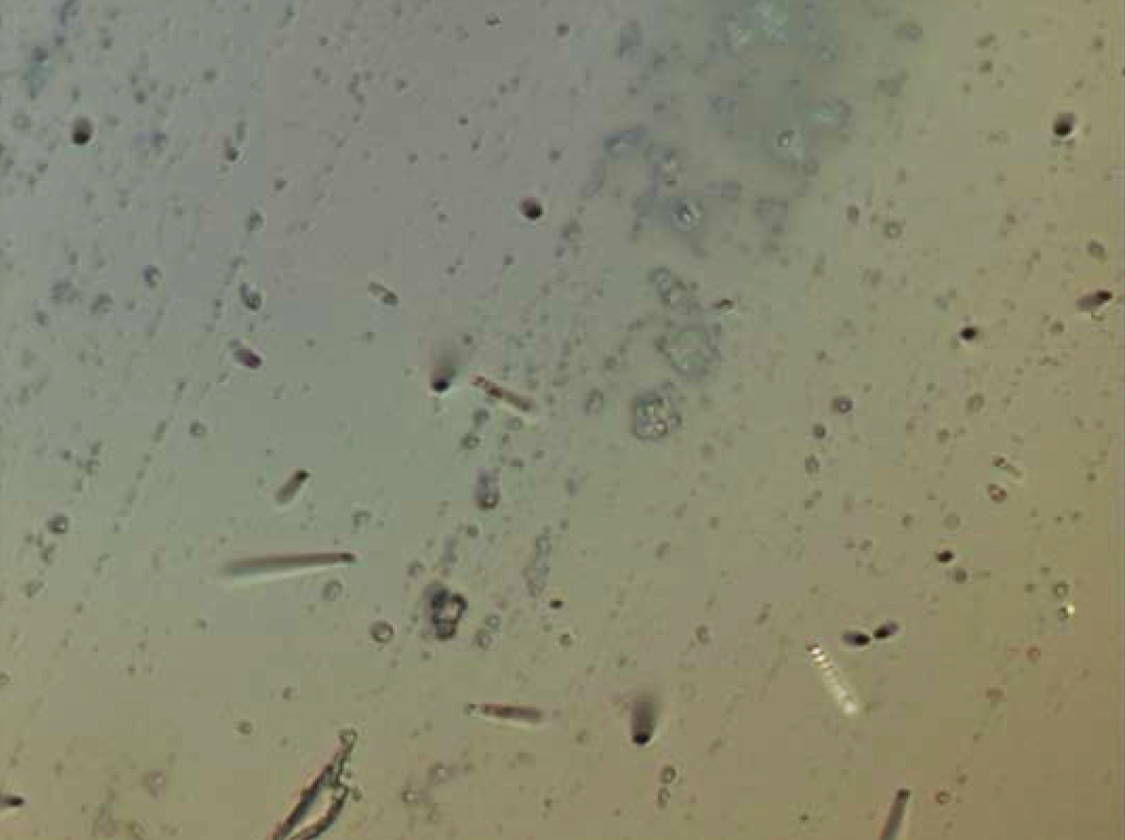}}}\\
{\mbox{\includegraphics[width=70mm]{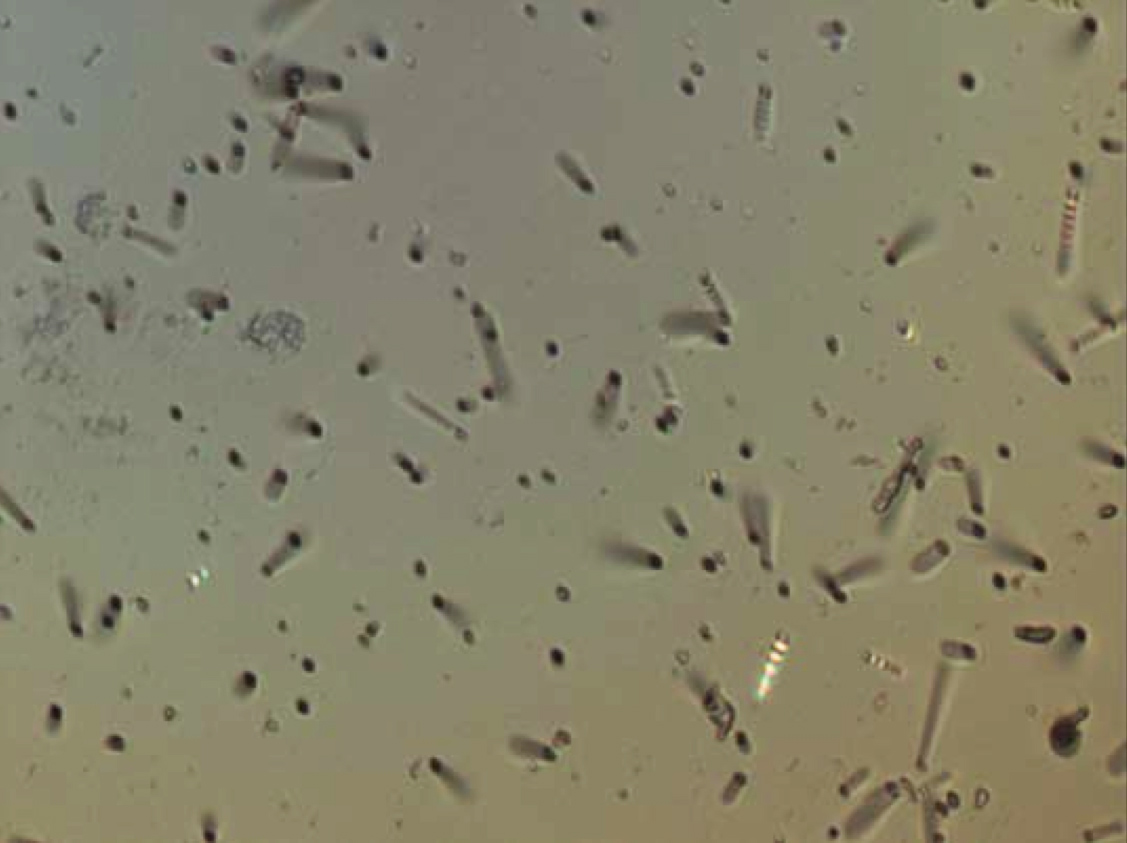}}}\\
{\mbox{\includegraphics[width=70mm]{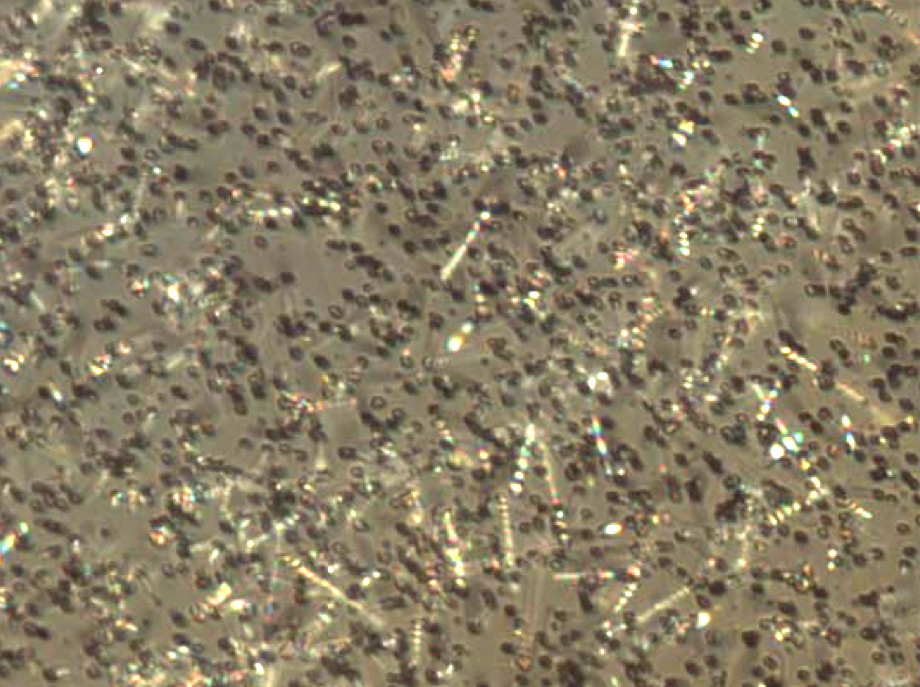}}}
   \end{tabular}
\end{center}
\caption{Reflected-light optical-microscopy images of fission tracks in the mica slides from Fig.~\ref{f-ALL}
 for the three irradiation fluences,
$\Phi^1_p$ (top), $\Phi^2_p$ (middle) and $\Phi^3_p$ (bottom).\label{f-REF}}
\end{figure}

Additional information can be obtained from a comparison of images obtained in transmitted
(Fig.~\ref{f-TRA}) and
reflected (Fig.~\ref{f-REF}) light. The two sets of images give consistent pictures.
In particular, no etched, confined tracks are visible, as one would observe, if present, through the occasional
intersection with surface tracks. Further, no tracks penetrate through the
mica, as revealed by etch pits being present at all track terminations: the source of the tracks is external to the mica.
From all these observations, the origin is clearly established to lie in the projection
of break-up fragments from nuclei in the Lead Tungstate that underwent fission.
\begin{table}[H]
\begin{center}
\begin{tabular}{|c|c|c|}
\hline
p fluence & track density  &  Ratio \\
~[$10^{12}$cm$^{-2}$] & [cm$^{-2}$]& ~[$10^{5}$]  \\
 \hline
   & & \\
$0.12\pm0.01$ & $ (2.9\pm 0.1)\times10^5 $& $4.0\pm 0.5$ \\
$1.59\pm0.13$ & $ (2.8\pm 0.1)\times10^6 $& $5.6\pm 0.5$ \\
$7.3\pm0.9$ & not counted & -  \\
   & & \\
\hline    
\end{tabular}
\end{center}
\caption{Densities of observed tracks counted on the mica slides. No accurate
count was possible at the highest fluence. }
\label{t-TB1}
\end{table}  

\section{Complementary tests on Lead Tungstate}
The methods developed in geochronology, to use mica as an external detector, 
allowed us to avoid having to etch the activated Lead Tungstate slides.
An observation of those slides was however attempted, without a surface treatment, in a Sigma scanning electron microscope from Zeiss~\cite{r-ZEI}  equipped with a HKL Advance electron backscattered diffraction system from Oxford Instruments~\cite{r-OXF}.
For each surface element hit by the electron beam during the scanning, an electron diffraction pattern is obtained and the local crystallographic orientation calculated from it. The quality of the pattern is influenced by a number of factors including local crystalline perfection, to account for it a quality value according to the contrast and sharpness of the diffraction pattern is also attributed to each surface element. Orientation maps and quality value maps provide a useful visualization of regions with different lattice orientation and with disturbed lattice structure respectively but is limited by the lateral resolution of the technique estimated to be of 200 nm (size of the smallest distinguishable details obtained in a scratched sample of the same material using the same analysis conditions).
In order to avoid any artefact resulting from lapping and polishing micro scratches, cleaved faces obtained by breaking irradiated and not-irradiated slides were examined. As expected from a single crystal free of lattice damage, the orientation and quality value in the non-irradiated sample were homogeneous throughout the mapped region. The same results were obtained from the irradiated sample.
This is not a surprise: typical fission track dimensions are small compared with the resolution of the technique.

All of these features are  compatible with the observation of Rayleigh
scattering in Lead Tungstate and of tracks due to fission fragments in mica, in that
\begin{enumerate}
\item[a)] the fission fragments damage the crystal through their extremely high energy loss
over a short path;
\item[b)] damage can be followed by an immediate recrystallization ~\cite{r-TOM};
\item[c)] optical boundaries, which act as scatterers, are due to mechanical strains remaining after re-crystallization~\cite{r-TOL}.
\end{enumerate}

It was in fact noticed e.g., by  D. L. Mills that 
Rayleigh-type light scattering can occur simply due to the damage regions having
``optical boundaries'' from mechanical strains after re-crystallization,
: {\em ``\ldots the inclusion of scattering from the strain
halo \ldots  leaves the well-known $\omega^4$ dependence unaltered''}~\cite{r-MIL}.

\begin{figure}[H]
\begin{center}
{\mbox{\includegraphics[width=120mm]{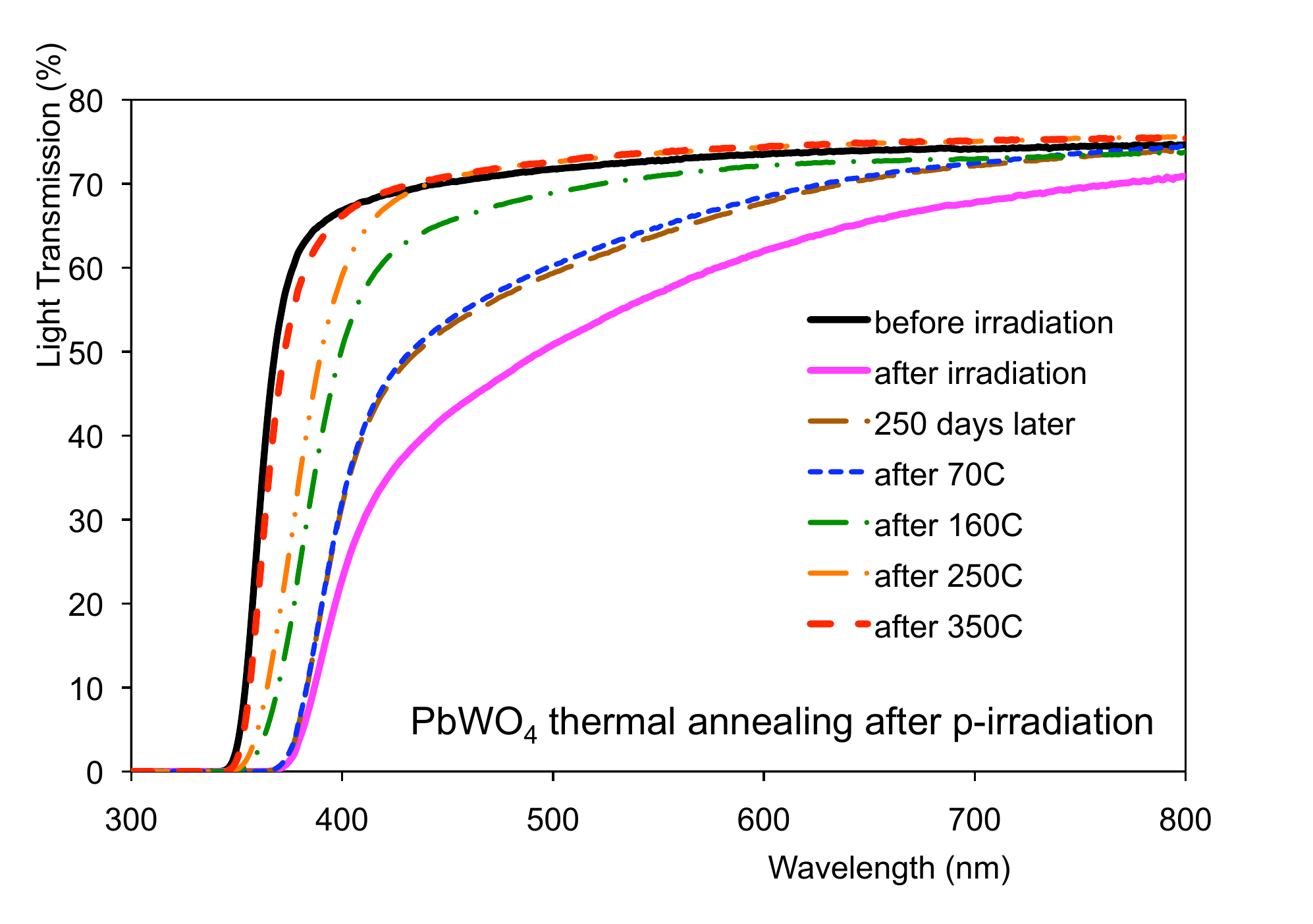}}}
\end{center}
\caption{Transmission curves for a proton irradiated crystal, 23 cm long,
after heating in air up to different temperatures }\label{f-DRE}
\end{figure}

It might be interesting to note that the scattering centers
can be eliminated by heating the crystals up to 350\,$^o$C~\cite{r-LTNIM}, and that 
such a heating treatment is also used after ingot production to
anneal mechanical stresses before cutting and polishing~\cite{r-ANN}. 
The progressive recovery of crystal light transmission through
heating up to different temperatures is visible in Fig.~\ref{f-DRE}, where transmission
curves are shown after the crystal was subjected
to various heating cycles  performed in an increasing temperature order.
The crystal, labelled ``d'' in ~\cite{r-LTNIM}, was annealed 240 days after
irradiation. The annealing temperature was reached in a programmable oven, 
by increasing the temperature linearly over 2 hours. The crystal was then kept
at the target annealing temperature for 8 hours. The temperature of the oven was
then reduced to room temperature over two hours.
While essentially no recovery of the crystal was
observed at 160\,$^o$C, partial recovery took place at
250\,$^o$C. At a temperature
of 350\,$^o$C almost complete recovery was reached, except for a small
residual shift of the band-edge. Shining LASER light through the
crystal after the damage annealing by heating, no scattering centers are
observed.
\section*{Conclusions}
Macroscopic observations show that light scattering centers are left in
Lead Tungstate after hadron irradiation, which are permanent at room temperature,
as described in our earlier work~\cite{r-LTNIM}.
The work presented here has allowed to visualize this damage at a microscopic scale,
as damage tracks originating in Lead Tungstate and entering mica slides
used as an external detector.
Scanning Electron Microscope imaging tests indicate the damage regions in
Lead Tungstate are in average crystalline but for - possibly - a track core, that is not visible
due to the finite spacial resolution of the technique.
It is understood that Rayleigh-type light scattering can occur, due to the damaged regions
having ``optical boundaries'' from mechanical strains remaining after re-crystallization.

The microscopic observations presented in this work confirm our understanding of the
damage mechanism due to hadrons in Lead Tungstate, where the fragments from
the fission induced in Lead and Tungsten cause severe, local damage to the
crystalline lattice due to their extremely high energy loss over a short track.
This evidence is relevant, in that it confirms our
understanding  of the damage mechanisms involved.

This confirmed understanding
will have to be taken into account  when selecting materials for high-energy physics 
calorimeters, where an exposure to
large integrated fluences of energetic hadrons is expected.
Specifically, we have confirmed that scintillators should be avoided, that are composed by elements
with a Z  $>$ 71, which is the threshold~\cite{r-THR} at which fission becomes important.
\section*{Acknowledgements}
We are indebted to the CERN PS accelerator team, that provided us with the required
 beam conditions for the proton irradiations. We are deeply
grateful to M.~Glaser, who operated the proton irradiation facility
and provided the Aluminium foil dosimetry.

\end{document}